\documentclass[12pt]{article}
\usepackage{amsmath,amsthm,amssymb}
\usepackage{color}
\usepackage{graphicx,subfigure,float}
\usepackage{enumitem}
\usepackage[round]{natbib}
\usepackage{geometry}
\usepackage{multirow}
\usepackage{booktabs}
\usepackage{placeins}
\usepackage{xr}
\def\argmin{\mathop{\rm argmin}\limits}

\newcommand{\mI}{\mathcal{I}}

\newcommand{\mC}{\mathcal{C}}

\newcommand{\mbX}{\boldsymbol{X}}
\newcommand{\mbx}{\boldsymbol{x}}
\newcommand{\bE}{\mathbb{E}}

\newcommand{\suit}[1]{\left( #1\right)}

\newcommand{\set}[1]{\left\{ #1\right\}}

\newcommand{\abs}[1]{\left| #1\right|}

\newcommand{\bolbeta}{\boldsymbol{\beta}}

\newcommand{\hatbolbeta}{\widehat{\bolbeta}}
\newcommand{\hatR}{\widehat{R}}

\newcommand{\hatgamma}{\widehat{\gamma}}

\newcommand{\floor}[1]{\lfloor #1 \rfloor}

\newcommand{\Names}{\textnormal{Max-Linear Tail Regression}}
\newcommand{\names}{\textnormal{max-linear tail regression }}
\theoremstyle{plain}
\newtheorem{theorem}{Theorem}
\newtheorem{corollary}{Corollary}
\newtheorem{prop}{Proposition}
\newtheorem{lemma}{{\bf Lemma}}

\theoremstyle{remark}
\newtheorem{remark}{Remark}

\author{
    \text{Liujun Chen}\thanks{International Institute of Finance, School of Management, University of Science and Technology of China,   ljchen22@ustc.edu.cn}, \
    \text{Deyuan Li}\thanks{ School of Management, deyuanli@fudan.edu.cn}
      \ and Zhengjun Zhang \thanks{School of Econimics and Management, University of Chinese Academy Sciences; Department of Statistics, University of Wiscosin,  zjz@stat.wisc.edu}
 }
\begin{document}
\title{ \Names}
\maketitle
 
\begin{abstract}
    The relationship between a response variable and its covariates can vary significantly, especially in scenarios where covariates take on extremely high or low values. This paper introduces a max-linear tail regression model specifically designed to capture such extreme relationships. To estimate the regression coefficients within this framework, we propose a novel M-estimator based on extreme value theory. The consistency and asymptotic normality of our proposed estimator are rigorously established under mild conditions.  Simulation results demonstrate that our estimation method outperforms the conditional least squares approach. We validate the practical applicability of our model through two case studies: one using financial data and the other using rainfall data.
\end{abstract}

{\it keywords: extreme value statistics, M-estimation, heavy tails, tail dependence}

\clearpage

\section{Introduction}

When covariates take on extremely high or low values, their effect on the response variable can differ significantly from their impact at typical levels \citep{teng2022two}. Understanding how covariates influence a response variable under these extreme conditions is critical in various fields, including environmental science, engineering, finance, economics, and social sciences. For example, evaluating the economic implications of extreme weather events or assessing the resilience of structures under severe loading conditions are vital tasks in these domains. Additionally, in financial risk management, analyzing extreme market movements is crucial for developing robust strategies to mitigate risks. Our research is motivated by this need to accurately model extreme behaviors, and while we focus on financial applications, the proposed models and methodologies can be easily adapted to other fields, such as studying rainfall amount or disaster preparedness in engineering.

In financial risk management, substantial attention is devoted to portfolio performance during adverse market conditions.
Empirical studies show  that asset returns exhibit a stronger correlation with  market returns during market  downturns, compared to their overall levels  \citep{longin1995correlation, longin2001extreme, ang2002asymmetric}.
Several models have been proposed to  evaluate the sensitivity of asset returns to market returns during adverse market conditions.  For instance,
\cite{ang2006downside} and \cite{lettau2014conditional} introduced the  downside risk capital asset pricing model  to estimate downside beta for  assets.
\cite{van2016systematic,van2019estimating} further extended the analysis to estimate the tail beta when the market index incurs substantial losses.


In analyzing a multivariate factor model during market crashes, two critical issues arise.   The first issue is defining extremely adverse market conditions.  In this paper, we focus on the scenario where the multivariate factor random vector $\mbX = (X_1,\dots,X_d) \in \mathbb{R}^d$ falls into a tail region
$$
\mC_0 = \set{\mbx: x_1>F_1^{-1}(1-p_0) \quad \text{or} \ \cdots \  \text{or} \quad x_d>F_d^{-1}(1-p_0)},
$$
where $0<p_0<1$ is a small constant and  $^\leftarrow$ denotes the left-continuous inverse function.  Here, $F_j$ denotes the marginal distribution of $X_j$, for $j=1,\dots,d$.   Such a tail region is commonly employed in the literature of multivariate extremes, see e.g. \cite{kiriliouk2022estimating} and \cite{huet2023regression}.

The second issue is developing an appropriate model to combine different risk factors. Traditional factor models and the downside risk factor models proposed by \cite{farago2018downside} typically employ linear functions to combine factors. However, given the extreme value context, a max-linear model may offer a more suitable framework. The max-linear structure has been applied in various contexts, including graphical models \citep{gissibl2018max, gissibl2021identifiability}, tail dependence modeling \citep{zhang2011asymptotic, zhang2017random}, factor models \citep{cui2018max}, and regression models \citep{cui2021max}.

In this paper, we propose a \names model for situations in which the factors $\mbX$ fall into the tail region $\mC_0$.  Mathematically, we model the  response variable $Y$ and the covariates $\mbX$  by
$$
Y = f_0(\mbX)\mI\suit{\mbX \in \mC_0}+ f_1(\mbX) \mI\suit{\mbX \notin \mC_0} +\varepsilon,
$$
where
 $f_0(\mbX) = \max(\beta_1X_1,\dots, \beta_dX_d) $ with  $\beta_1,\dots,\beta_d>0$ and  $\varepsilon$ represents random noise.  Throughout the paper,  we do not make any parametric assumption on $f_1(\mbX)$ as  our focus is on  the tail regions.   {
 Focusing on a subsample of the dataset is a powerful strategy in modern statistical analysis, allowing for the investigation of specific subpopulations or regions of interest that are critical for addressing targeted research questions.  For example, \cite{heckman1979sample} demonstrated the importance of accounting for selection bias in subsampled data, highlighting its role in producing more accurate and reliable inferences. }

 Our model functions as a threshold model, distinguishing the behavior  of the response $Y$
 based on whether the covariates fall into specific regions. However, unlike traditional threshold models \citep{hansen2000sample, hansen2017regression, teng2022two}, our model specifically addresses the situation when
$\mbX$ falls into a tail region, i.e., $p_0$ is small.
Studying extreme events is valuable,  as noted by statistician John  Tukey:
``As I am sure almost every geophysicist knows, distributions of actual errors and fluctuations have much more straggling extreme values than would correspond to the magic bell-shaped distribution of Gauss and Laplace''.



 Estimating   the coefficients $\beta_1,\dots, \beta_d$ in our \names model  presents substantial challenges.   Typically,  coefficients of threshold models can be estimated using conditional least squares  approach \citep{lettau2014conditional,teng2022two}.
 However, in the context of tail regression, where $p_0$
  is small, the sample size within
$\mC_0$ is limited, which can lead to high variance in estimates, rendering the traditional conditional least squares approach less reliable. Moreover, this approach is not robust to heavy-tailed data,  which is
often the characteristic of financial returns \citep{zhao2018modeling}.  Heavy-tailed distributions are also prevalent in other domains, such as environmental science \citep{buishand2008spatial} and gene expression \citep{ke2022high}.

To address these challenges, we propose alternative estimators based on extreme value analysis.
 Extreme value analysis, which focuses on the statistical inference on the
 tail of a distribution, has  proven to be an effective tool for handling extreme events. We refer interested readers to the monographs \cite{beirlant2004statistics}, \cite{haan2006extreme}
 and \cite{resnick2007heavy} for a general introduction of extreme value analysis.
   Employing a max-linear tail  regression model with  heavy-tailed covariates, we first  study the tail behavior of the response variable $Y$ and investigate  the tail dependence structure between $Y$ and $\mbX$. We then introduce an M-estimator for the coefficient vector $\bolbeta$, establishing its consistency and asymptotic normality under mild conditions. Simulation results demonstrate that the extreme value analysis based estimator performs better than the conditional least squares estimator.

 Our model  relates to the work of     \cite{cui2021max}, but differs in that we assume a max-linear structure only in the tail region, rather than globally.  Additionally, our model generalizes the tail regression model proposed by \cite{van2016systematic, van2019estimating}, which focused on a univariate covariate. Extending this model to a multivariate setting is non-trivial due to the complexity of managing dependencies among covariates and the max-linear structure.
Furthermore, our model is distinct from the tail quantile regression model \citep{chernozhukov2011inference,wang2012estimation}. While tail quantile regression model links the conditional tail quantile of $Y$ to a linear combination of $\mbX$, our model predicts $Y$ using a max-linear combination of $\mbX$ conditional on  $\mbX$ falling into a tail region.


\section{Methodology}\label{sec:Methodology}

For the ease of presentation, we focus on the case $d=2$ in this section.  The results can be easily extended to the case $d\ge 3$, see Section \ref{sec:multi} for details.  We consider the   max-linear tail regression model
\begin{equation}\label{eq:model}
    Y = \max(\beta_1 X_1,\beta_2X_2)  +\varepsilon, \ \text{for} \ \set{X_1>F_1^{\leftarrow}(1-p_0) \ \text{or}\  X_2>F_2^{\leftarrow}(1-p_0)}.
\end{equation}
Recall that we do not impose any parametric assumption on
$f_1(\mbX)$ and consequently on
 $Y$ for $\set{X_1\le F_1^{\leftarrow}(1-p_0), X_2\le F_2^{\leftarrow}(1-p_0)}$.

 \subsection{Identification}

 We begin by describing the heavy-tailedness of $X_1$ and $X_2$ within the framework of  extreme value theory. Suppose that there exist constants $\gamma_j>0, j=1,2$ such that for all $x>0$,
\begin{equation}\label{eq:heavy-tail-x}
        \lim_{t\to\infty}\frac{U_j(tx)}{ U_j(t)}=   x^{\gamma_j}, \quad j=1,2,
\end{equation}
where $U_j(t) = F_j^{\leftarrow}(1-1/t)$. The parameter $\gamma_j$,   known as the \textit{extreme  value index}, governs  the tail behavior of $X_j$.
 To quantify the speed of convergence in \eqref{eq:heavy-tail-x}, we usually assume the second order conditions, i.e.,
there exist regular varying functions $A_j$ with index $\rho_j<0$, $j=1,2$, such that, as $t\to\infty$,
\begin{equation}\label{soc:rv:x}
        \sup_{x>1}\abs{x^{-\gamma} \frac{U_j(tx)}{U_j(t)}-1}=O\set{A_j(t)}.
\end{equation}
These  conditions are commonly used in extreme value statistics, see e.g. \cite{de1996generalized} and Chapter 3 of \cite{haan2006extreme}.


To investigate the tail behavior of $Y$ under model \eqref{eq:model}, it is crucial to model the tail dependence structure of  $(X_1,X_2)$. Assume that, for all $(x_1,x_2)\in (0,\infty)^2$, the following limit exists:
\begin{equation}\label{eq:tail:dependence:X}
    \lim_{p\downarrow 0}\frac{1}{p} \Pr\set{X_1> U_1\suit{\frac{1}{px_1}}, X_2>U_2\suit{\frac{1}{px_2}}  } = R^X(x_1,x_2).
\end{equation}
The function $R^X$, also known as the upper tail copula   \citep{schmidt2006non}, captures the tail dependence between $X_1$ and $X_2$.

Without loss of generality, we assume that $\gamma_1\ge \gamma_2$. Under the tail dependence model \eqref{eq:tail:dependence:X},
the  extreme value index of
$\max(\beta_1 X_1,\beta_1 X_2)$ is equal to $\gamma_1$.  We further assume that the noise term $\varepsilon$  has a thinner tail compared to  $\max(\beta_1 X_1,\beta_1 X_2)$.  Mathematically,
there exits some $\delta>0$, such that
\begin{equation}\label{eq:epsilon}
    \bE |\varepsilon|^{\frac{1+\delta}{\gamma_1}}<\infty.
\end{equation}
Let us highlight that no independence assumption is made between $(X_1,X_2)$ and $\varepsilon$.  Additionally, it is not assumed that $\varepsilon$ has a mean of zero.

The relation \eqref{eq:model} is specified only for the tail region of $(X_1,X_2)$. Let $F_{L}$ denote the distributions of $Y$ conditional on $\set{X_1 \leq F_1^{\leftarrow}(1-p_0), X_2 \leq F_2^{\leftarrow}(1-p_0)}$.  We do not impose any parameter assumption on $F_L$; instead, we assume that, $F_L$ is continuous and  has a tail that is not heavier than that of   $X_1$. Mathematically, we assume that,   there exists  a constant $c_L\ge 0$ such that,
\begin{equation}\label{eq:model:Central}
    \frac{U_L(t)}{U_1(t)} \to c_L,
\end{equation}
as $t\to\infty$,
where
$U_L(t) = F_L^{\leftarrow}(1-1/t)$.

Under these conditions, we  establish the tail behavior of $Y$  in the following proposition.
\begin{prop}\label{theorem:identification}
Assume that the model \eqref{eq:model} and conditions \eqref{soc:rv:x}-\eqref{eq:model:Central} hold.
If $\gamma_1 = \gamma_2$, then there exist constants $\alpha_1>0,\alpha_2>0$,
such that, as $t\to\infty$,
$$
\frac{U_Y(t)}{U_1(t)} \to \alpha_1,\quad \frac{U_Y(t)}{U_2(t)} \to \alpha_2,
$$
where $U_Y(t) = F_Y^{\leftarrow}(1-1/t)$. If $\gamma_1 >\gamma_2$, there exists a constant $\alpha_0>0$, which does not depend on  $\beta_2$,  such that, as $t\to\infty$,
$$
\frac{U_Y(t)}{U_1(t)} \to \alpha_0.
$$
\end{prop}

Proposition \ref{theorem:identification} implies that,   if $\gamma_1>\gamma_2$,
 the high quantile of $Y$ is  exclusively influenced   by  $X_1$.  {Thus, the parameter $\beta_2$ is not
 identifiable from the first order tail behavior of $(X_1,X_2,Y)$.
 }
  To ensure identifiability of the model parameters,
 we assume that $\gamma_1 = \gamma_2=:\gamma$
  for the rest of the paper. {Such an equal extreme value index assumption
  is foundational to several theoretical models, for example, the multivariate regular variation model  \citep{resnick2007heavy, mainik2015portfolio}. Moreover, this assumption is also adopted by spatial extremes models \citep{buishand2008spatial, einmahl2022spatial}.
  This assumption will be justified for our real data application.  In practice, if this assumption does not hold, marginal transformations can be applied to ensure its validity.}

  Next, we explore the tail dependence structure between $(X_1,X_2)$ and $Y$.
Denote
$$
R_{p}(x_1,x_2,y) = \frac{1}{p}\Pr\set{X_1>U_1\suit{\frac{1}{px_1}},X_2>U_2\suit{\frac{1}{px_2}}, Y>U_Y\suit{\frac{1}{py}}},
$$
where $(x_1,x_2,y)\in (0,\infty]^3\backslash \set{(\infty,\infty,\infty)}$. We establish the limit of $R_{p}(x_1,x_2,y)$  in the following theorem.

\begin{theorem}\label{theorem:prob}
    Assume that the  conditions  in Proposition \ref{theorem:identification} hold with $\gamma_1=\gamma_2=\gamma$. Then, for any $(x_1,x_2,y)\in (0,\infty]^3\backslash \set{(\infty,\infty,\infty)}$,
     as $p\downarrow 0$,
$$
\begin{aligned}
    R_p(x_1,x_2,y) \to& R(x_1,x_2,y) \\
    := &R^X\suit{ \min\set{  (\beta_1/\alpha_1)^{1/\gamma}y, x_1 },x_2} +R^X\suit{   x_1, \min\set{ (\beta_2/\alpha_2)^{1/\gamma}y, x_2 }} \\
    &- R^X\suit{\min\set{  (\beta_1/\alpha_1)^{1/\gamma}y, x_1 }, \min\set{ (\beta_2/\alpha_2)^{1/\gamma}y, x_2 }}.
\end{aligned}
$$
\end{theorem}
\begin{remark}
    Theorem 1 and Proposition 1 are established under the condition (\ref{eq:epsilon}).
However, if  $\varepsilon$ is independent of $(X_1,X_2)$, then condition (\ref{eq:epsilon}) can be relaxed to
    $$
       \lim_{t\to\infty} \frac{U_{\varepsilon}(t)}{U_1(t)} = c_{\varepsilon},
    $$
    where $U_{\varepsilon}(t) = F_{\varepsilon}^{\leftarrow}(1-1/t)$ and $c_{\varepsilon} \ge 0$.  This relaxation
    allows for the tail of $\varepsilon$ to be as heavy as that of $X_1$ (and $X_2$).
    The corresponding theorems and their proofs are provided in the Supplementary Material.
\end{remark}

By applying Theorem \ref{theorem:prob} with $(x_1,x_2,y)=(1,\infty,1)$ and $(x_1,x_2,y)=(\infty,1,1)$, we derive the following corollary, which forms the  basis of our estimation  methodology for $\bolbeta = (\beta_1,\beta_2)$.

\begin{corollary}\label{theorem:prob:corollary}
 Assume the same conditions as in Theorem \ref{theorem:identification}. Then,
$$
\begin{aligned}
    R(1,\infty,1)= & \theta_1 + R\suit{1,\theta_2,\infty} - R(\theta_1,\theta_2,\infty), \\
    R(\infty,1,1)= & \theta_2 + R\suit{\theta_1,1,\infty} - R(\theta_1,\theta_2,\infty), \\
\end{aligned}
$$
where
$$
\begin{aligned}
  \theta_1 =   \suit{\frac{\alpha_1}{\beta_1}}^{-1/\gamma},\quad
  \theta_2 =   \suit{\frac{\alpha_2}{\beta_2}}^{-1/\gamma}.
\end{aligned}
$$
Moreover, we have that, $\theta_1< 1$ and $\theta_2< 1$.
\end{corollary}
\begin{remark}
    If $X_1$ and $X_2$ are tail independent, i.e., $R^X(x,y) =0$ for all $(x,y)\in (0,\infty)^2$, then Corollary \ref{theorem:prob:corollary} simplifies to
    $$
    \begin{aligned}
        \beta_1 = & \suit{R(1,\infty,1)}^{\gamma} \alpha_1, \quad \text{and} \quad
        \beta_2 = & \suit{R(\infty,1,1)}^{\gamma} \alpha_2. \\
    \end{aligned}
    $$
    This result aligns with the findings on tail regression with  a univariate covariate, see Theorem 1 of \cite{van2019estimating}.
\end{remark}

\subsection{Estimation}
In this subsection, we propose an M-estimator for $\bolbeta$ based on Corollary \ref{theorem:prob:corollary}.  Our data consists of independent  copies $\set{X_{i1},X_{i2}, Y_i}_{i=1}^n$ of $(X_1,X_2,Y)$.
First, we estimate
the function $R$, the parameters $\alpha_1, \alpha_2$ and $\gamma$ based on extreme value analysis. Then, by solving  the equations in Corollary \ref{theorem:prob:corollary}, we can obtain the estimates for $\bolbeta$.  For simplicity, we consider an intermediate sequence $k $ satisfying $k=k(n)\to\infty, k/n\to 0$ as $n\to\infty$,
 in the estimation process.

We start with the estimation of the $R$ function. Let $r_{i,j}$ denote the rank of $X_{ij}$  among $X_{1j},\dots,X_{nj}$, $i=1,\dots,n$, $j=1,2$, and  $r_{i,Y}$ denote the rank of $Y_i$ among $Y_1,\dots,Y_n$, $i=1,\dots,n$.
Define
$$
\hatR(x,y,z)=\frac{1}{k}\sum_{i=1}^n \mI \suit{r_{i,1}>n-kx+\frac{1}{2}, \ r_{i,2}>n-ky+\frac{1}{2}, \ r_{i,Y}>n-kz+\frac{1}{2}},
$$
see e.g. \cite{drees1998best} and  \cite{einmahl2012m}.
The constant $1/2$ in the argument of the indicator function
helps to improve the finite-sample properties of the estimator.

{
Next, we  estimate $\alpha_1$ and $\alpha_2$.
Proposition \ref{theorem:identification} states that,
$$
\alpha_1=\lim_{p\downarrow 0} \frac{U_Y(1/p)}{U_1(1/p)}.
$$
This suggests  the estimator
$$
\widehat{\alpha}_1= \int_{s_0}^1  \frac{Y_{(n-\floor{ks})}}{X_{(n-\floor{ks}),1}}ds,
$$
where $s_0\in (0,1)$,  $Y_{(1)}\le Y_{(2)}\le \cdots Y_{(n)}$ are the order statistics of $Y_1,\dots, Y_n$ and $X_{(1),1}\le X_{(2),1}\dots\le X_{(n),1}$ are the order statistics of $X_{11}, \dots, X_{n1}$.
 Similarly we estimate $\alpha_2$ by
$$
\widehat{\alpha}_2=\int_{s_0}^1  \frac{Y_{(n-\floor{ks})}}{X_{(n-\floor{ks}),2}}ds,
$$
where $X_{(1),2}\le X_{(2),2}\dots\le X_{(n),2}$ are the order statistics of $X_{12}, \dots, X_{n2}$.
For the following numerical studies, we select $s_0=1/2$. Other choices of $s_0$ are also examined, and these variations do not significantly impact the performance of our estimators.
}

We proceed with the estimation of the extreme value index $\gamma$. Note that, $X_1$, $X_2$ and $Y$ all follow heavy-tailed distributions with extreme value index $\gamma$. We propose using the estimator
$$
\hatgamma = \frac{1}{3}\suit{\hatgamma_1+\hatgamma_2+\hatgamma_Y},
$$
where $\hatgamma_1, \hatgamma_2$ and $\hatgamma_Y$ are the Hill estimators \citep{hill1975simple} based on the samples of $X_1,X_2$ and $Y$, respectively. More specifically,
$$
\begin{aligned}
    \hatgamma_1= \frac{1}{k}\sum_{i=1}^k \log\frac{X_{(n-i+1),1}}{X_{(n-k),1}}, \quad
    \hatgamma_2= \frac{1}{k}\sum_{i=1}^k \log\frac{X_{(n-i+1),2}}{X_{(n-k),2}}, \quad
    \hatgamma_Y= \frac{1}{k}\sum_{i=1}^k \log\frac{Y_{(n-i+1)}}{Y_{(n-k)}}.\\
\end{aligned}
$$

Finally, we  define $\hatbolbeta =(\widehat{\beta}_1$, $\widehat{\beta}_2)$ as the solution to the following equations,
$$
\begin{aligned}
    \hatR(1,\infty,1)
     =&   \suit{\widehat{\alpha}_1/\widehat{\beta_1} }^{-1/\hatgamma} +\widehat{R}\suit{1, \suit{\widehat{\alpha}_2/\widehat{\beta_2} }^{-1/\hatgamma},\infty}
    - \hatR\suit{\suit{\widehat{\alpha}_1/\widehat{\beta_1} }^{-1/\hatgamma} ,\suit{\widehat{\alpha}_2/\widehat{\beta_2} }^{-1/\hatgamma},\infty}, \\
    \hatR(\infty,1,1) =&    \suit{\widehat{\alpha}_2/\widehat{\beta_2} }^{-1/\hatgamma} +\widehat{R}\suit{ \suit{\widehat{\alpha}_1/\widehat{\beta_1} }^{-1/\hatgamma},1,\infty}
    - \hatR\suit{\suit{\widehat{\alpha}_1/\widehat{\beta_1} }^{-1/\hatgamma} ,\suit{\widehat{\alpha}_2/\widehat{\beta_2} }^{-1/\hatgamma},\infty}.
\end{aligned}
$$

\subsection{Asymptotic theory}\label{sec:theory}
In this subsection, we investigate the asymptotic theories of $\widehat{\bolbeta}$.
We assume that the function $R^X$ has first order partial derivatives
$$
\begin{aligned}
    R_1^X(x_1,x_2) = \frac{\partial R^X(x_1,x_2)}{\partial x_1}, \quad
    R_2^X(x_1,x_2) = \frac{\partial R^X(x_1,x_2)}{\partial x_2}.
\end{aligned}
$$
The partial derivative functions
$R_1^X(x_1,x_2)$ and $ R_2^X(x_1,x_2)$ exhibit the following properties.
\begin{lemma}[\cite{schmidt2006non}] \label{lemma:R:property}
    For all $x_1,x_2>0$,
    $$
    0\le R_1^X(x_1,x_2) \le 1, \quad 0\le R_2^X(x_1,x_2) \le 1.
    $$
    Moreover,  the functions $x_1\mapsto R_2^X(x_1,x_2)$ and $x_2\mapsto R_1^X(x_1,x_2)$ are non-decreasing.
\end{lemma}

To establish the asymptotic theories of $\widehat{\bolbeta}$, we need to
 make a slightly stronger condition on $R^X_1$ and $R^X_2$:
\begin{equation}\label{eq:condition:tech:dirivates}
    \begin{aligned}
        R^X_1\suit{\theta_1,\theta_2}<1,& \quad
        R^X_2\suit{\theta_1,\theta_2}<1,  \\
        R^X_1\suit{\theta_1,\theta_2}< R^X_1\suit{\theta_1,1},& \quad
        R^X_2\suit{\theta_1,\theta_2}< R^X_2\suit{1,\theta_2}, \\
    \end{aligned}
\end{equation}
where $\theta_1$ and $\theta_2$ are defined in Corollary \ref{theorem:prob:corollary}.
Condition \eqref{eq:condition:tech:dirivates} is imposed   to ensure that $\beta_1,\beta_2$ are the  ``well-separated'' zeros  of the population estimation equations in Corollary \ref{theorem:prob:corollary}. We establish the consistency of $\widehat{\bolbeta}$  in the following theorem.
\begin{theorem}\label{theorem:consistency}
    Assume the same conditions as in Theorem \ref{theorem:prob}. Moreover, assume that  condition \eqref{eq:condition:tech:dirivates} holds and the parameter space $\boldsymbol{\Theta}$ is compact.
  Then, as $n\to\infty$,
    $$
    \begin{aligned}
         \hatbolbeta\stackrel{P}{\to} \bolbeta.
    \end{aligned}
    $$
\end{theorem}

Next, we deal with  the asymptotic normality of our estimator. To proceed,
we  require a second order condition for the distribution function of $Y$.
By the proof of Proposition \ref{theorem:identification},
the second order condition of the distribution function of  $Y$ is driven by the second order conditions of the distributions of  $X_1,X_2, \varepsilon$ and  $F_{L}$. Thus, without loss of generality, we  directly impose a second order condition on the distribution function of $Y$, that is, there exists a  regular varying function $A_Y$ with index $\rho_Y<0$,  such that, as $t\to\infty$,
\begin{equation}\label{soc:rv:y}
        \sup_{x>1}\abs{x^{-\gamma} \frac{U_Y(tx)}{U_Y(t)}-1}=O\set{A_Y(t)}.
\end{equation}

Additionally, we require  second order conditions to quantify the speed of convergence of the tail dependence function $R$. Similarly, the second order condition of $R$ can be derived by combining the second order conditions of $R^X, U_1,U_2, F_{L}$ and $F_{\varepsilon}$. Thus, without loss of generality,  we directly assume the second order conditions on the $R$ function. Specifically, we assume that, for some $\tau>0$,
as $p\downarrow 0$,
\begin{equation}\label{eq:SOC:tail:dependence}
    \begin{aligned}
       R_{p}(x_1,x_2,y)- R(x_1,x_2,y) =& O(p^{\tau}),\\
       R_{p}(\infty,x_2,y)- R(\infty,x_2,y) =& O(p^{\tau}),\\
       R_{p}(x_1,\infty,y)- R(x_1,\infty,y)=& O(p^{\tau}),\\
       R_{p}(x_1,x_2,\infty)- R(x_1,x_2,\infty) =& O(p^{\tau}), \\
    \end{aligned}
\end{equation}
uniformly for $(x_1,x_2,y)\in (0,T]^3$ with $T>0$.

Finally, we assume that the intermediate sequence $k$ satisfies that, as $n\to\infty$,   $k\to\infty$ and $k=O(n^{\eta})$ where
\begin{equation}\label{eq:condition:k}
    0<\eta<\min \suit{\frac{2\tau}{1+2\tau}, \frac{-2\rho^*}{1-2\rho^*} },
\end{equation}
with $\rho^*=\max(\rho_1,\rho_2,\rho_Y)$.  The following theorem provides the asymptotic normality of $\widehat{\bolbeta}$.
\begin{theorem}\label{theorem:normality}
    Assume that the conditions in Theorem \ref{theorem:consistency} hold.  Moreover, assume that conditions \eqref{soc:rv:y}, \eqref{eq:SOC:tail:dependence} and \eqref{eq:condition:k} hold.
     Then, as $n\to\infty$,
    $$
    \begin{aligned}
        \sqrt{k} \suit{\hatbolbeta -\bolbeta}\stackrel{d}{\to} N\left( \boldsymbol{0}, \boldsymbol{\Sigma}\right).
    \end{aligned}
    $$
    The covariance structure $\boldsymbol{\Sigma}$  is defined  in the Supplementary Material.
\end{theorem}

{
    Our asymptotic theories, i.e.,  Theorems \ref{theorem:consistency} and \ref{theorem:normality}, are derived under the assumption of independent and identically distributed (i.i.d.) data. While this assumption simplifies the mathematical derivation, the results can be extended to stationary time series, which often provide a more realistic representation of real-world data exhibiting serial dependence.

    A closer inspection of the proofs for Theorems \ref{theorem:consistency} and \ref{theorem:normality} reveals that the i.i.d. assumption is primarily used to establish the asymptotic properties of the estimators for the tail dependence function $\widehat{R}$, as well as for the  tail quantities $\widehat{\alpha}_1$, $\widehat{\alpha}_2$, and $\widehat{\gamma}$. For stationary time series, these properties can still be established under appropriate conditions on the serial dependence structure of the data \citep{drees2000weighted,drees2010limit}. By applying the same reasoning used in the proofs of Theorems \ref{theorem:consistency} and \ref{theorem:normality}, we can  establish the asymptotic theories of $\hatbolbeta$ under the assumption of stationary time series.
    A detailed discussion of this extension to stationary time series is provided in   the Supplementary Material.
}





\section{Extension to high dimensions}\label{sec:multi}
In this  section, we extend our results to the case   $d\ge 3$. The \names model assumes
$$
Y = \max(\beta_1X_1,\dots,\beta_pX_p)+\varepsilon, $$
for  $\set{X_1>F_1^{-1}(1-p_0) \ \text{or} \cdots \text{or}\ X_d>F_d^{-1}(1-p_0)}$.

Assume that $U_j(t) = F_j^{-1}(1-1/t)$, $ j=1,\dots,d$  satisfy the second order conditions \eqref{soc:rv:x} with same extreme value index $\gamma$. Assume that the tail dependence function $R^X$ exists, i.e., for $(x_1,\dots,x_d) \in (0,\infty)^d$,
$$
R^X(x_1,\dots,x_d):  =   \lim_{p\downarrow 0}\frac{1}{p} \Pr\set{X_1> U_1\suit{\frac{1}{px_1}},\dots, X_d>U_d\suit{\frac{1}{px_d}}  }.
$$

For any sequence $S= \set{i_1,\dots,i_J}$, where $i_j\in \set{1,\dots,n}$ and $i_1\ne \cdots\ne i_J$, and a vector $(x_{1},\dots,x_{J}) \in (0,\infty)^J$,  we define
$$
R^{S}(x_{i_1},\dots,x_{i_J}):=R^X(\mbx),
$$
where $\mbx$ is a $d$-dimensional vector with $i_j$-th element being $x_{j}$, $j=1,\dots,J$ and other elements being $\infty$.
For instance, for any $(x_1,x_2,x_3) \in (0,\infty)^3$,
$$
R^{\set{3,1,4}}(x_1,x_2,x_3)=R^X(x_2,\infty,x_1,x_3,\infty,\dots,\infty).
$$
Under the similar conditions  specified in Section  \ref{sec:Methodology}, we can show that,
for $j=1,\dots,d$,
$$
\begin{aligned}
    &\widetilde{R}^{\suit{j}}(1,1) \\
    :=   &\lim_{p\downarrow 0}\frac{1}{p} \Pr\set{X_j> U_j\suit{\frac{1}{p}}, Y>U_Y\suit{\frac{1}{p}}  }  \\
    =
    & \theta_j +\sum_{i\ne j} R^{\set{i,j}}(\theta_i,1) \\
                    &-\suit{ \sum_{i\ne j} R^{\set{i,j}}(\theta_i,\theta_j)+\sum_{i_1 \ne i_2\ne j} R^{\set{i_1,i_2,j}}(\theta_{i_1}, \theta_{i_2},1) } \\
                    &+ \suit{ \sum_{i_1\ne i_2\ne j} R^{\set{i_1,i_2,j}}(\theta_{i_1},\theta_{i_2},\theta_j)+\sum_{i_1 \ne i_2 \ne i_3 \ne j} R^{\set{i_1,i_2,i_3,j}}(\theta_{i1}, \theta_{i2},\theta_{i3}, 1) }\\
                    &\cdots \\
                    &+(-1)^{d-2} \suit{ \sum_{i_1 \ne \dots \ne i_{d-2}\ne j}R^{\set{i_1,\dots, i_{d-2},j}}(\theta_{i_1},\dots,\theta_{i_{d-2}},\theta_j) +\sum_{i_1 \ne \dots \ne i_{d-1}\ne j}R^{\set{i_1,\dots, i_{d-1},j}}(\theta_{i_1},\dots,\theta_{i_{d-1}},1)  } \\
                    &+(-1)^{d-1} \sum_{i_1 \ne \dots \ne i_{d-1}\ne j}R^{\set{i_1,\dots, i_{d-1},j}}(\theta_{i_1},\dots,\theta_{i_{d-1}},\theta_j),
\end{aligned}
$$
where
$$
\theta_j = \suit{\frac{\alpha_j}{\beta_j}}^{-1/\gamma}\quad \text{and} \quad \alpha_j = \lim_{t \to \infty} \frac{U_Y(t)}{U_j(t)}.
$$
We provide the formulas for the case $d=3$ in the Supplementary Material.

Then, by estimating $\gamma,\alpha_j$, $R^X(x_1,\dots,x_d)$ and $ \widetilde{R}^{\suit{j}}(1,1) $, we can estimate $\beta_1,\dots, \beta_d$ by solving the $d$ estimation equations. The consistency and the asymptotic normality of the estimators can be established in a similar manner   as for the case  $d=2$.

\section{Simulation}\label{sec:simu}

\subsection{Simulation  for the case $d=2$}
In this section, we conduct a simulation study to investigate the finite sample performance of the proposed estimator. We compare our proposed estimator with the conditional least squares estimator $(\widetilde{\beta}_0, \dots, \widetilde{\beta}_d)$, which is defined as
$$
\argmin_{\beta_0,\dots,\beta_d} \sum_{i=1}^n (Y_i - \beta_0 - \max(\beta_1X_{i1},\dots,\beta_dX_{id}) )^2\mI\suit{  X_{i1} \ge X_{(n-k),1} \ \text{or} \ \cdots  \ \text{or} \  X_{id} \ge X_{(n-k),d}}.
$$

The first simulation study is conducted for the case $d=2$.  The response $Y$ is generated by a segmented model
$$
Y = \left\{
    \begin{array}{ll}
         \max(\beta_1 X_1, \beta_2 X_2) +\varepsilon, & \ \text{if} \  \max(X_1,X_2)>t_{\nu, 0.98}, \\
         \beta_2X_1+\varepsilon, & \ \text{if} \max(X_1,X_2)\le \ t_{\nu, 0.98},
    \end{array}
\right.
$$
where $\beta_1 = 0.6, \beta_2 = 0.4$ and
$t_{\nu, 0.98}$ is the 98\% quantile of a univariate $t$ distribution with degree of freedom $\nu$.   We emphasize that the parameter structure for the case $\set{\max(X_1,X_2)\le \ t_{\nu, 0.98}}$ is  not utilized during the estimation process. We consider three different settings for the covariates $(X_1,X_2)$ and the random noise $\varepsilon$.

\begin{enumerate}
    \item[(M1)] The covariates
    $
    (X_1,X_2) \stackrel{i.i.d.}{\sim} t_{\nu}(\boldsymbol{S}),
    $
    with $$
    \boldsymbol{S} = \left[
        \begin{array}{cc}
            1 & 0.5\\
            0.5 & 1
        \end{array}
     \right].
    $$
    Here, $t_{\nu}(\boldsymbol{S})$ means a multivariate Student-t distribution with degree of freedom $\nu$ and scale matrix $\mathcal{S}$. The noise $\varepsilon  \stackrel{i.i.d.}{\sim} t_{\nu+1}$ and   is independent from $(X_1,X_2)$.

\item[(M2)]     The covariates $(X_1,X_2)$ are generated as in model (M1).  The noise variable  $\varepsilon = \frac{1}{3}\sqrt{|X_1|}+\sqrt{|X_2|}\varepsilon_0$, where $\varepsilon_0  \stackrel{i.i.d.}{\sim} t_{\nu+1}$    and is independent from $(X_1,X_2)$.
{
\item[(M3)]  $X_{i,1} =0.5X_{i-1,1} + e_{i,1}$ and $X_{i,2} = 0.5X_{i-1,2}+ e_{i,2}$. The innovations
$(e_{i,1}, e_{i,2}) \stackrel{i.i.d.}{\sim} t_{\nu}(\boldsymbol{S})$.   The noise $\varepsilon  \stackrel{i.i.d.}{\sim} t_{\nu+1}$ and   is independent from $(X_1,X_2)$.
}
\end{enumerate}

\begin{figure}[ht]
	\centering
	\subfigure[(M1), $\beta_1$]{
	\includegraphics[width=0.305\textwidth]{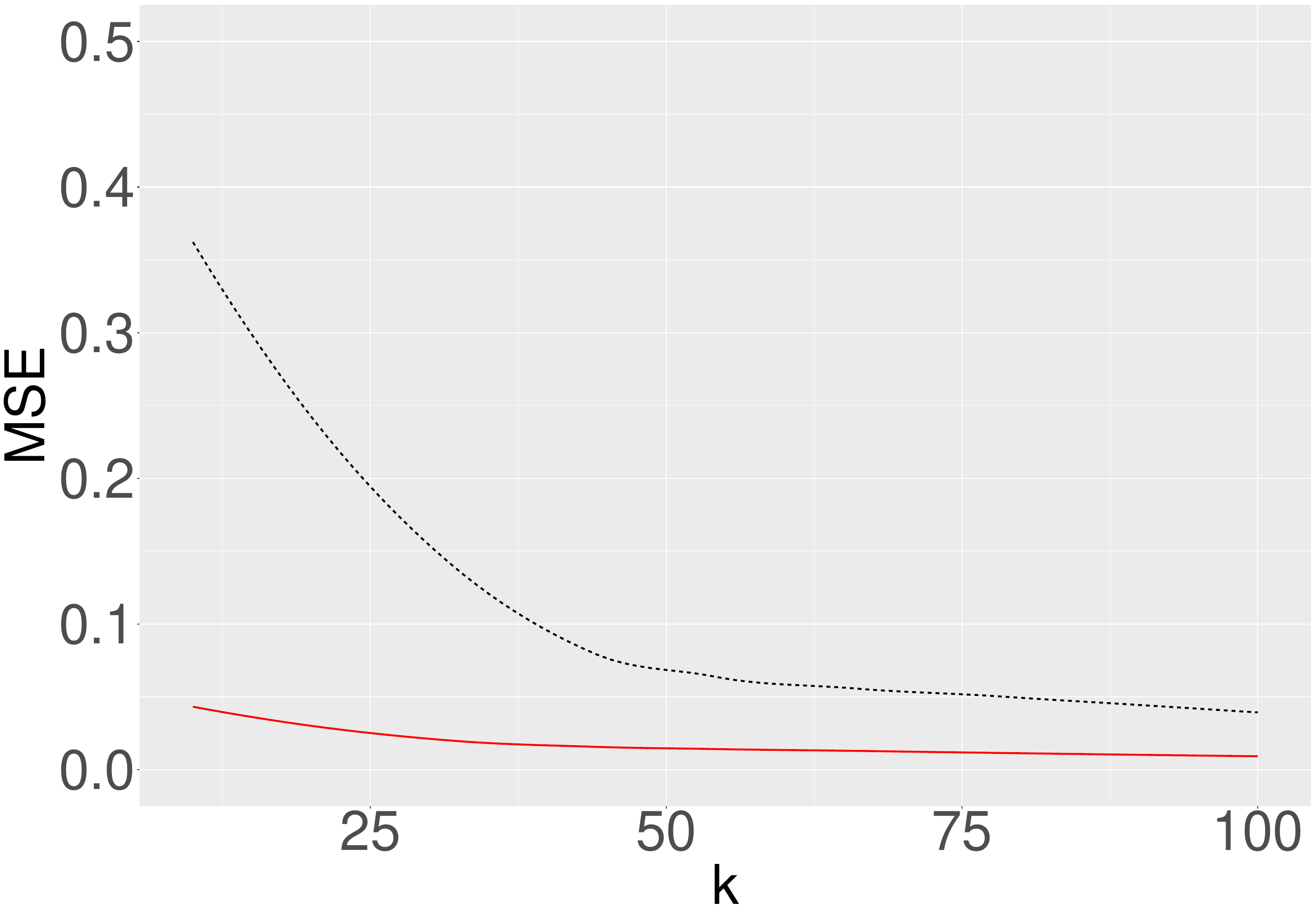}
	}
    \subfigure[(M2), $\beta_1$]{
        \includegraphics[width=0.305\textwidth]{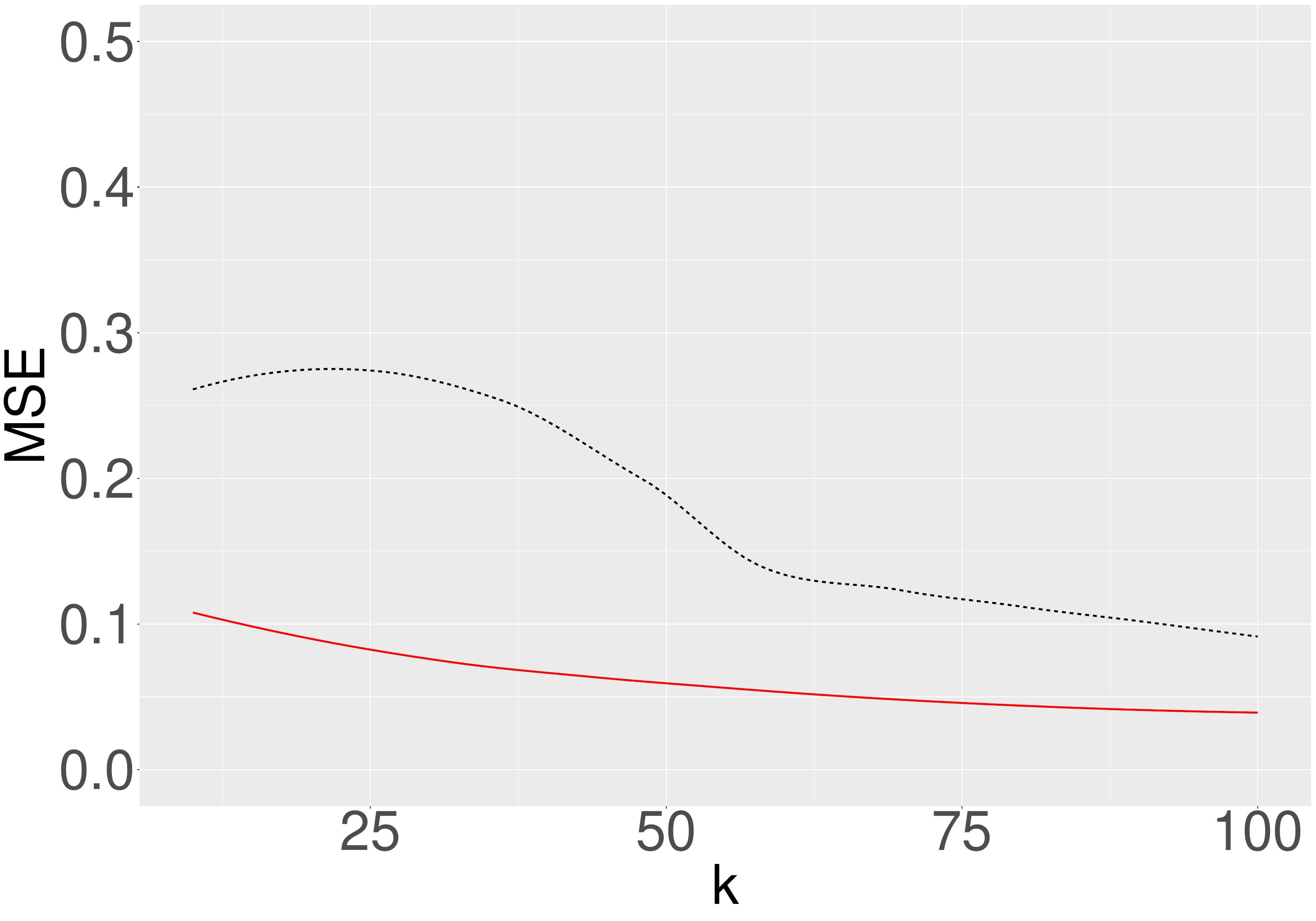}
    }
    \subfigure[(M3), $\beta_1$]{
        \includegraphics[width=0.305\textwidth]{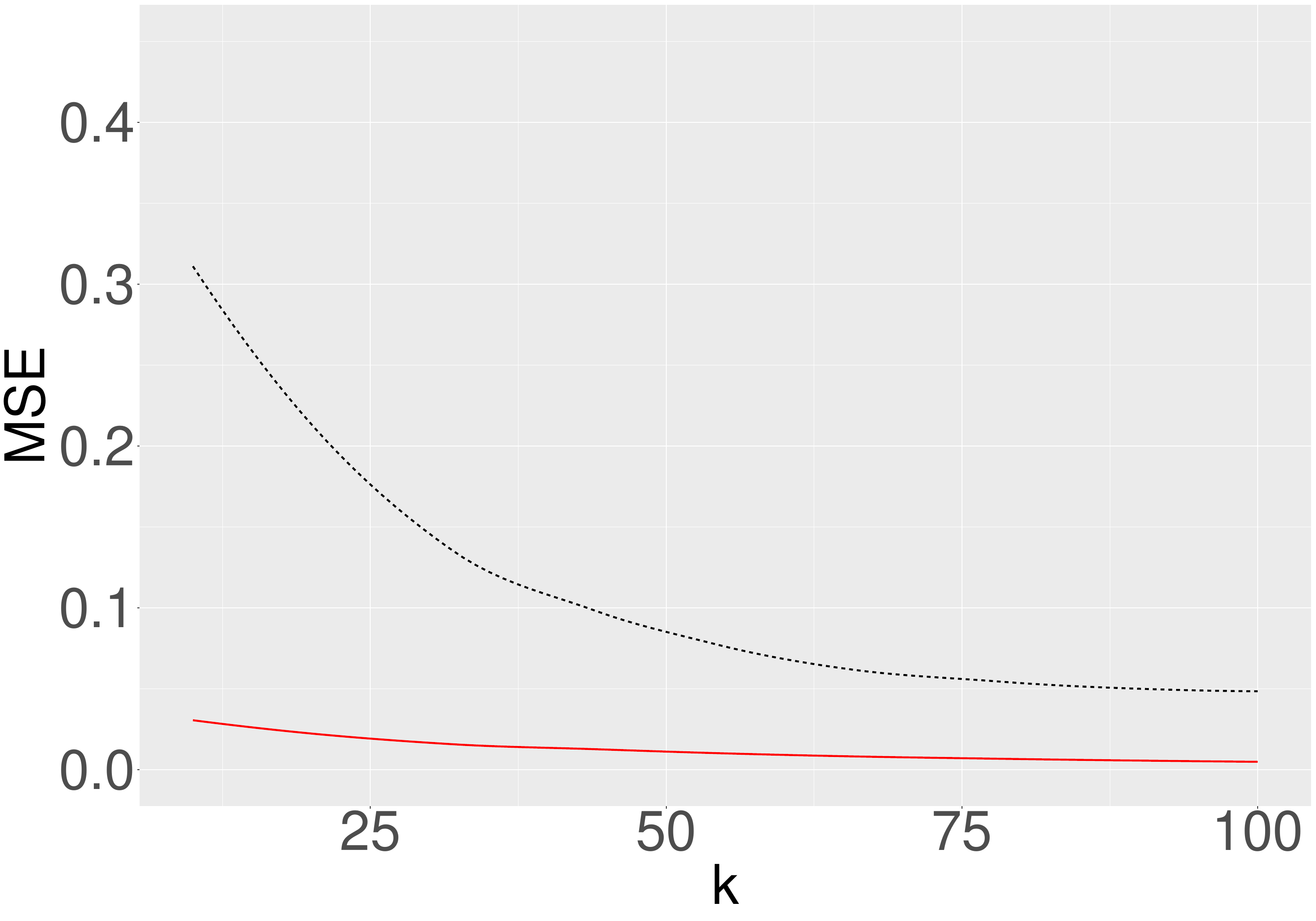}
    }
    \subfigure[(M1), $\beta_2$]{
        \includegraphics[width=0.305\textwidth]{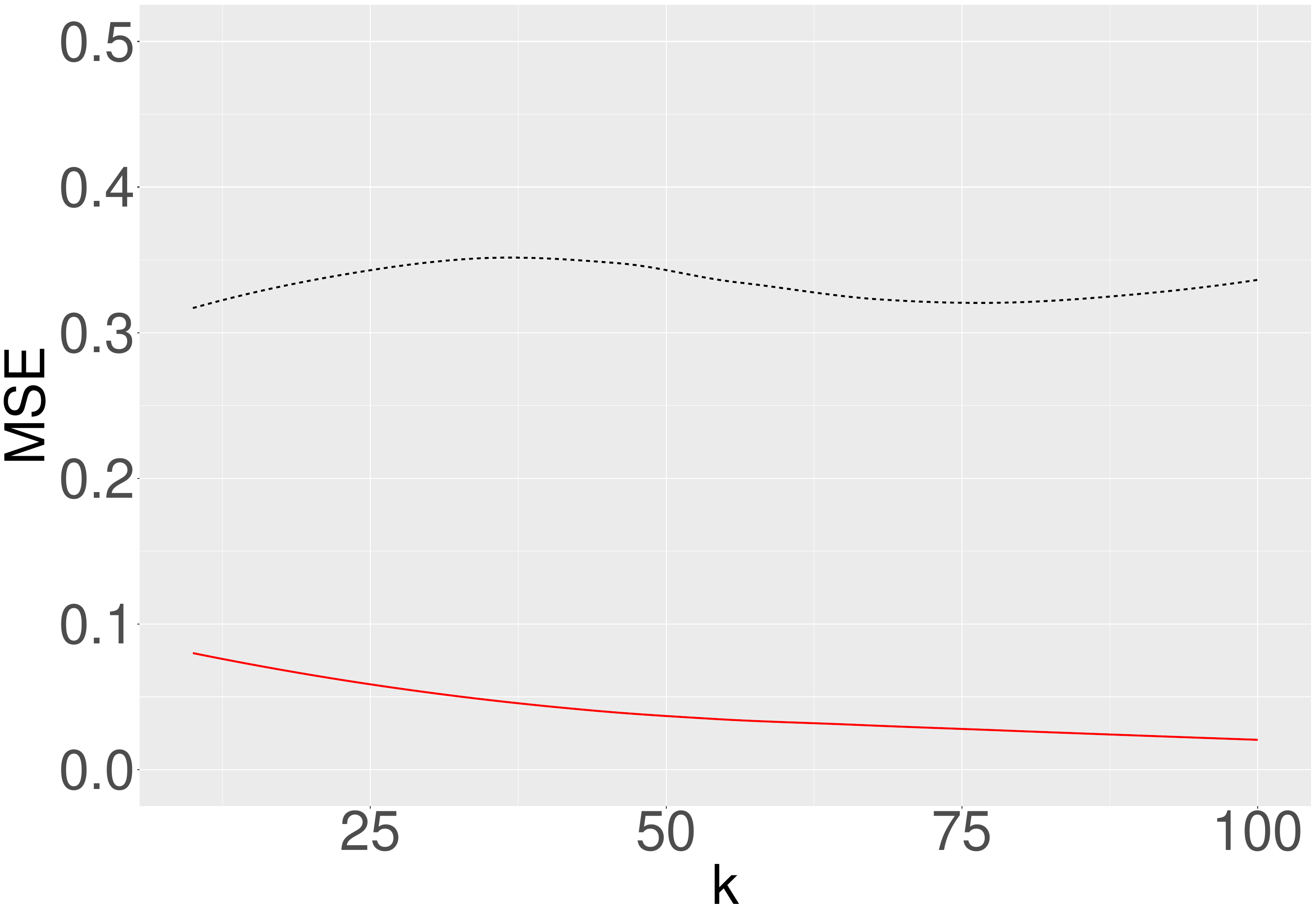}
    }
    \subfigure[(M2), $\beta_2$]{
        \includegraphics[width=0.305\textwidth]{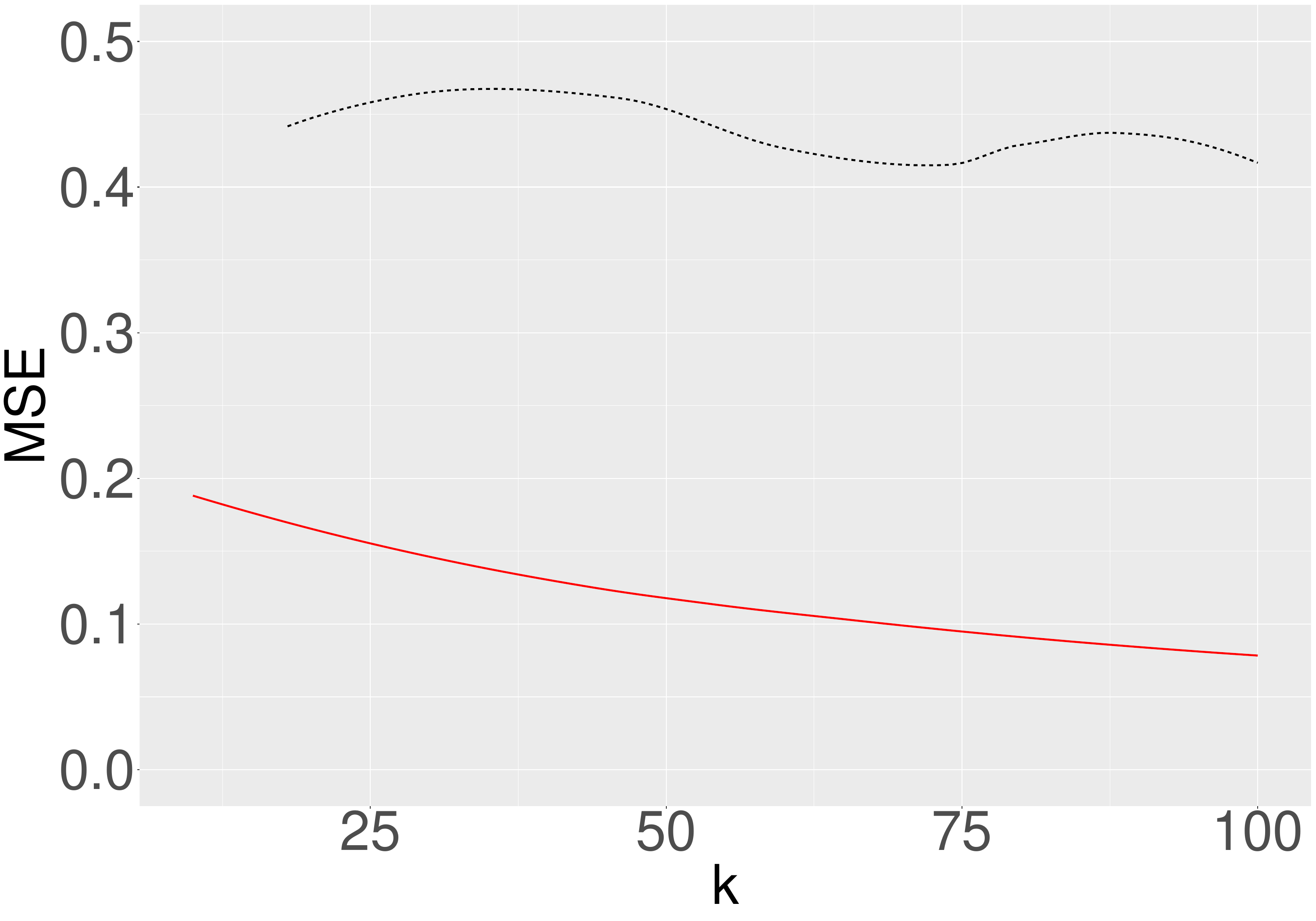}
    }
    \subfigure[(M3), $\beta_2$]{
        \includegraphics[width=0.305\textwidth]{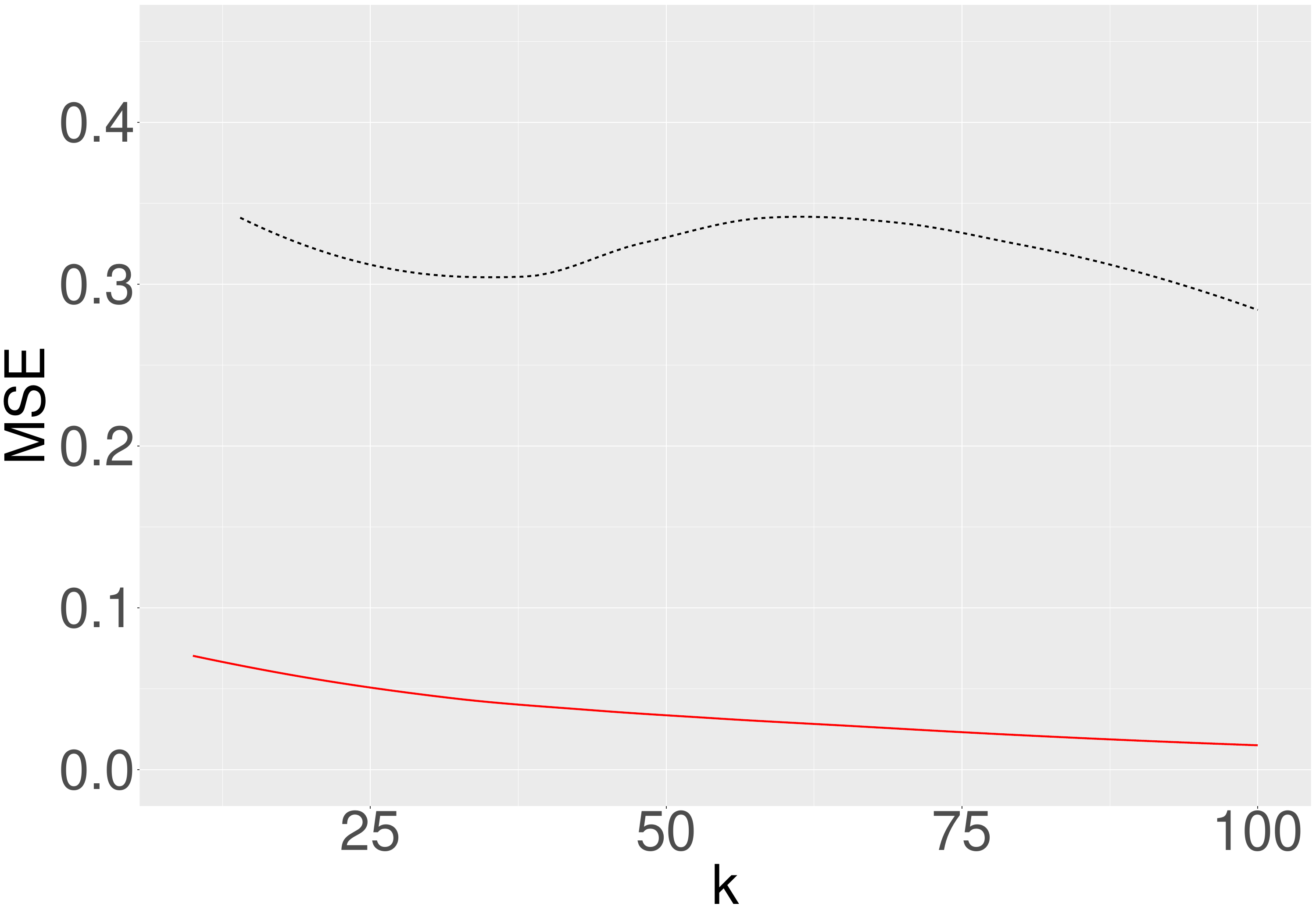}
    }
    \caption{MSE of the  estimators $\widehat{\beta}_j$({\textendash \textendash\textendash}) and $\widetilde{\beta}_j$({\color{black}$---$}) for  models (M1) and (M2) with  $\nu=4$.
    }
	\label{figure:mse:nu4}
	\end{figure}

In model (M3),  the covariates $(X_1,X_2)$ are generated through autoregressive processes.
We explore three different choices of $\nu$: $\nu = 3, 4$ and $5$.
For each  model, we generate $200$ samples with sample size $n=1000$ and estimate $\bolbeta$ in each sample by using the proposed estimator and the conditional least squares estimator. The mean squared error (MSE) of the two estimators are calculated. We  present the MSE across different values of $k$  with $\nu=4$ in Figure \ref{figure:mse:nu4}. The results for $\nu=3$ and $\nu=5$ are similar and have been omitted for brevity.

The proposed estimator demonstrates superior performance compared to the conditional least square estimator across all levels of $k$ from $10$ to $100$. In additional  to its enhanced performance, the proposed estimator exhibits greater robustness to the choice of $k$  and is less sensitive to heteroscedastic and dependent noise. {Furthermore, our proposed estimators still demonstrate strong performance when applied to time series data.}


\subsection{Simulation  for the case $d=3$}
In this subsection, we conduct a simulation study  for the case $d=3$.
The covariates $(X_1,X_2)$ is also generated from a bivariate $t$ distribution with degree of freedom $\nu$ and scale matrix $\boldsymbol{S}$.
The covariate $X_3\sim t_{v}$  and is independent from $(X_1,X_2)$.
The response $Y$ is   generated by a segmented model
$$
Y = \left\{
    \begin{array}{ll}
         \max(\beta_1 X_1, \beta_2 X_2, \beta_3X_3) +\varepsilon, & \ \text{if} \  \max(X_1,X_2, X_3)>t_{\nu, 0.98}, \\
         (\beta_1+\beta_2)X_3+\varepsilon, & \ \text{if} \  \max(X_1,X_2, X_3)\le t_{\nu, 0.98}, ,
    \end{array}
\right.
$$
where $\beta_1 = 0.4$, $\beta_2 = 0.3$ and $\beta_3 = 0.5$.   We generate $\varepsilon $ independently from  $t_{\nu+1}$. We focus exclusively on the independent noise setting because the choice between independent and  heteroscedastic noise does not significantly affect the outcomes. The sample size $n$ is fixed at $n = 1000$.

\begin{figure}[ht]
	\centering
	\subfigure[$\beta_1$]{
	\includegraphics[width=0.305\textwidth]{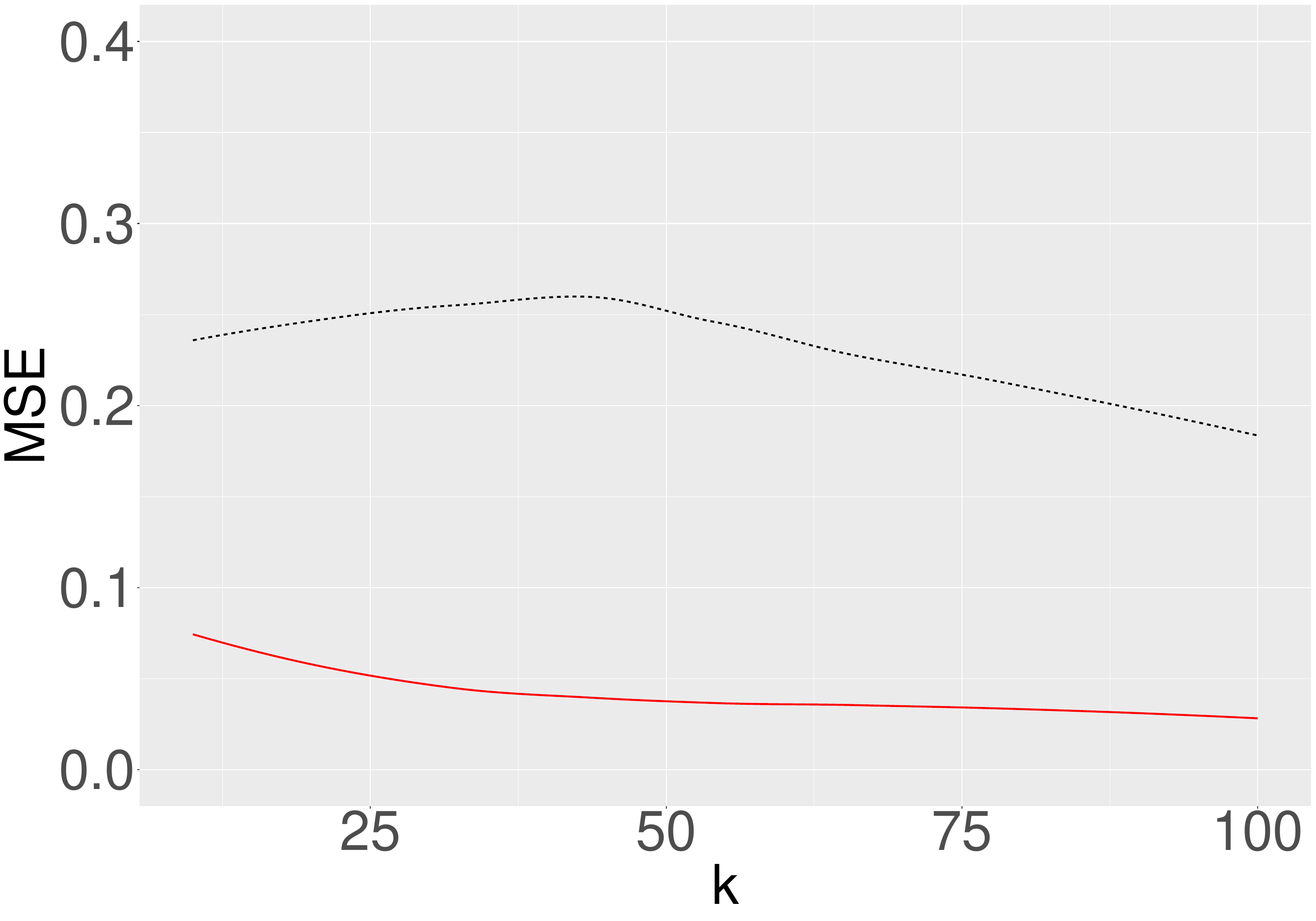}
	}
    \subfigure[$\beta_2$]{
        \includegraphics[width=0.305\textwidth]{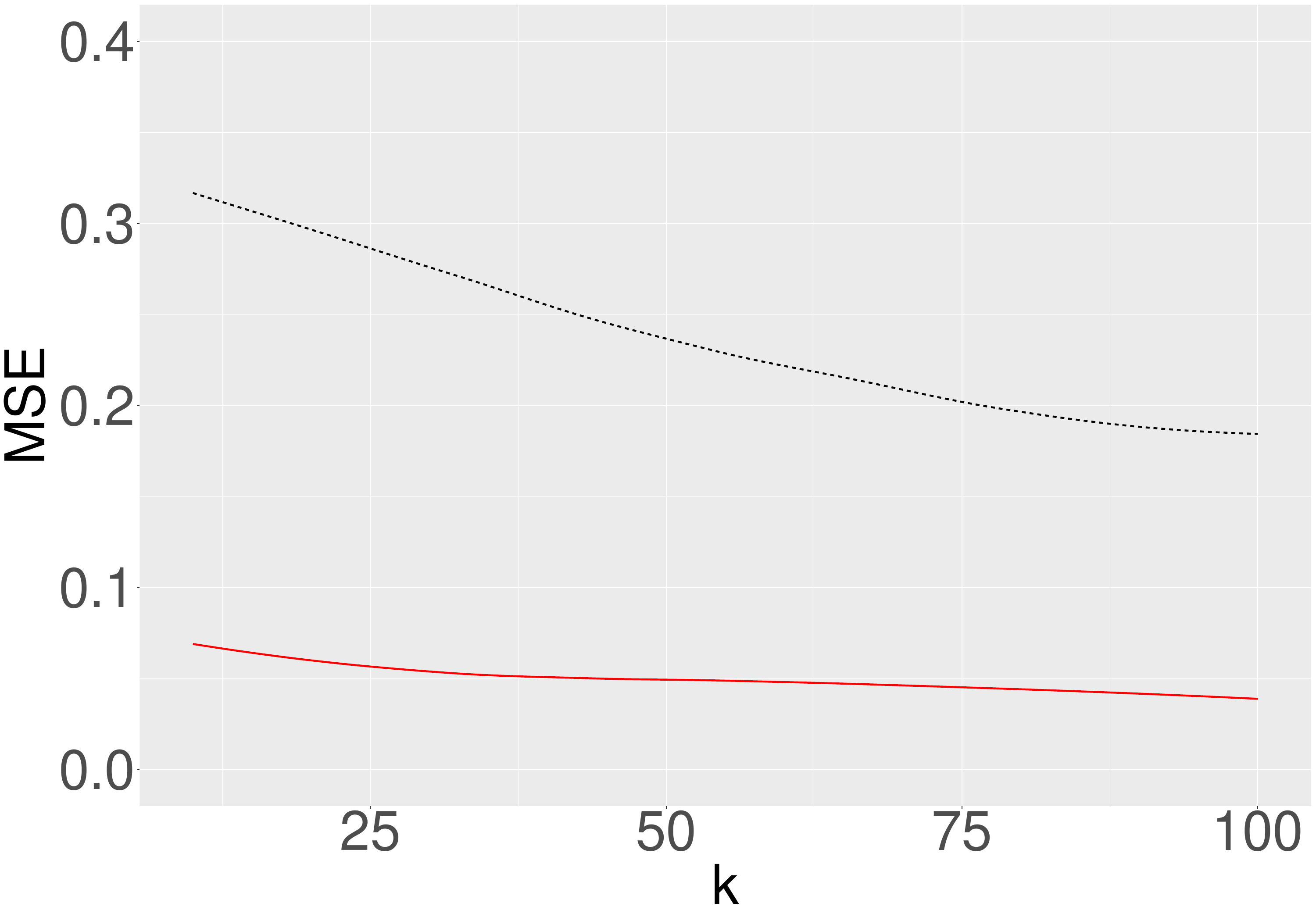}
    }
    \subfigure[$\beta_3$]{
        \includegraphics[width=0.305\textwidth]{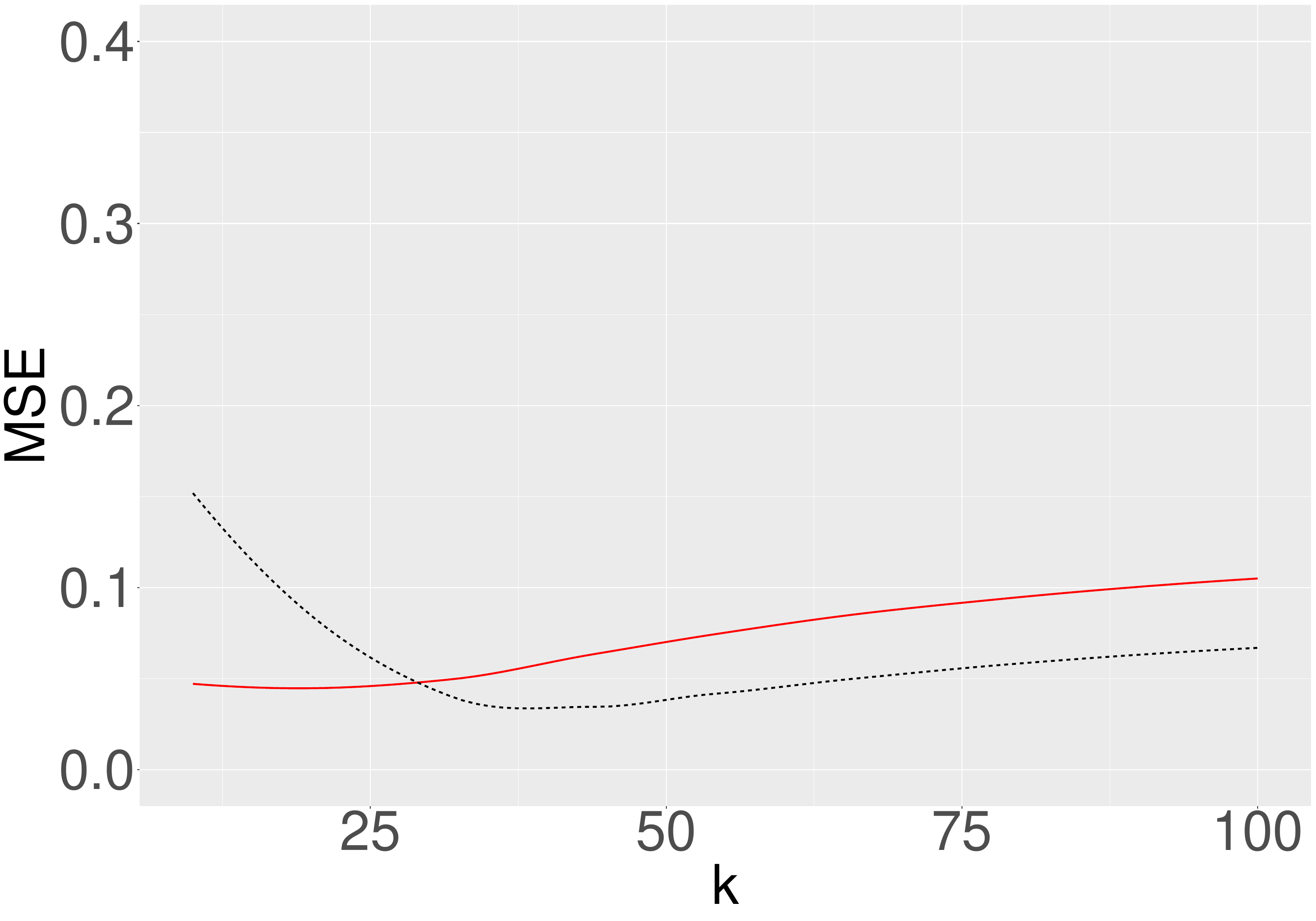}
    }
    \caption{MSE of the  estimators $\widehat{\beta}_j$({\textendash\textendash\textendash}) and $\widetilde{\beta}_j$({\color{black}$---$}) for the case $d=3$  with $\nu=4$.}
	\label{figure:mse:d3:nu4}
	\end{figure}

We consider three different choices for $\nu$: $\nu=3, 4$ and $5$. However, we only report the MSE for $\nu=4$ in Figure \ref{figure:mse:d3:nu4} since the results for $\nu=3$ and $5$ are similar. The observed patterns are consistent with those found in the case  $d=2$.
 Our proposed estimator outperforms the conditional least square approach for both $\beta_1$ and $\beta_2$ across all levels of $k$ from $10$ to $100$. For $\beta_3$, the  conditional least square estimator  slightly outperforms the proposed estimator for high levels of $k$.

\section{Application}\label{sec:application}
Risk pervades various systems, including economic, financial and environmental science. An important question in modern risk management is how to quantify the vulnerability of individual institutions to extreme system movements. In this section, we consider two representative system indices as covariates and apply the max-linear tail regression model to investigate the effects of systemic failures on individual entities.
\subsection{Financial market}\label{finance}

The first  application focuses on the financial market, where  we select $X_1$ as the  market index and $X_2$ as the cross-sectional maximum loss within the market.  This choice of $X_2$ is motivated by
a growing interest in the investigation of cross-sectional maximum loss within the market \citep{zhao2018modeling, zhang2021studying}.   We refer $X_1$ and $X_2$ to be the systematic index and systemic index of the system, respectively.

This analysis focuses on components of the Dow Jones Industrial Average (DJIA) spanning from January 1, 2000, to December 31, 2009, a period that includes the Great Recession.   The dataset comprises
$n=2514$ daily observations. We designate $X_1$ and $X_2$ to represent the negative log returns of the market index and
the maximum cross-sectional loss, respectively.
Additionally, let $Y$ denote the negative log return of an individual company. For a comprehensive overview of the entities involved, please refer to   the Supplementary Material.

{
Before applying our estimation procedures, we first validate the assumption that $\gamma_1 =\gamma_2 = \gamma_Y$.  To achieve this purpose, we  construct an confidence interval for $\gamma_1$ and check whether the estimators of $\gamma_2$ and $\gamma_Y$ fall within this interval.
Given that the financial data exhibits serial dependence, a confidence interval based on normal approximation using i.i.d. data may not be accurate. Therefore, we employ a declustering procedure to construct the confidence interval. Specifically, we take every second observation and treat the declustered observations as independent. The resulting confidence interval is $D = (0.194,0.411)$ (with $k=30$ for the declustered observations using a Hill plot).
}

\begin{figure}[htb]
    \centering
    \includegraphics[width=0.8\textwidth]{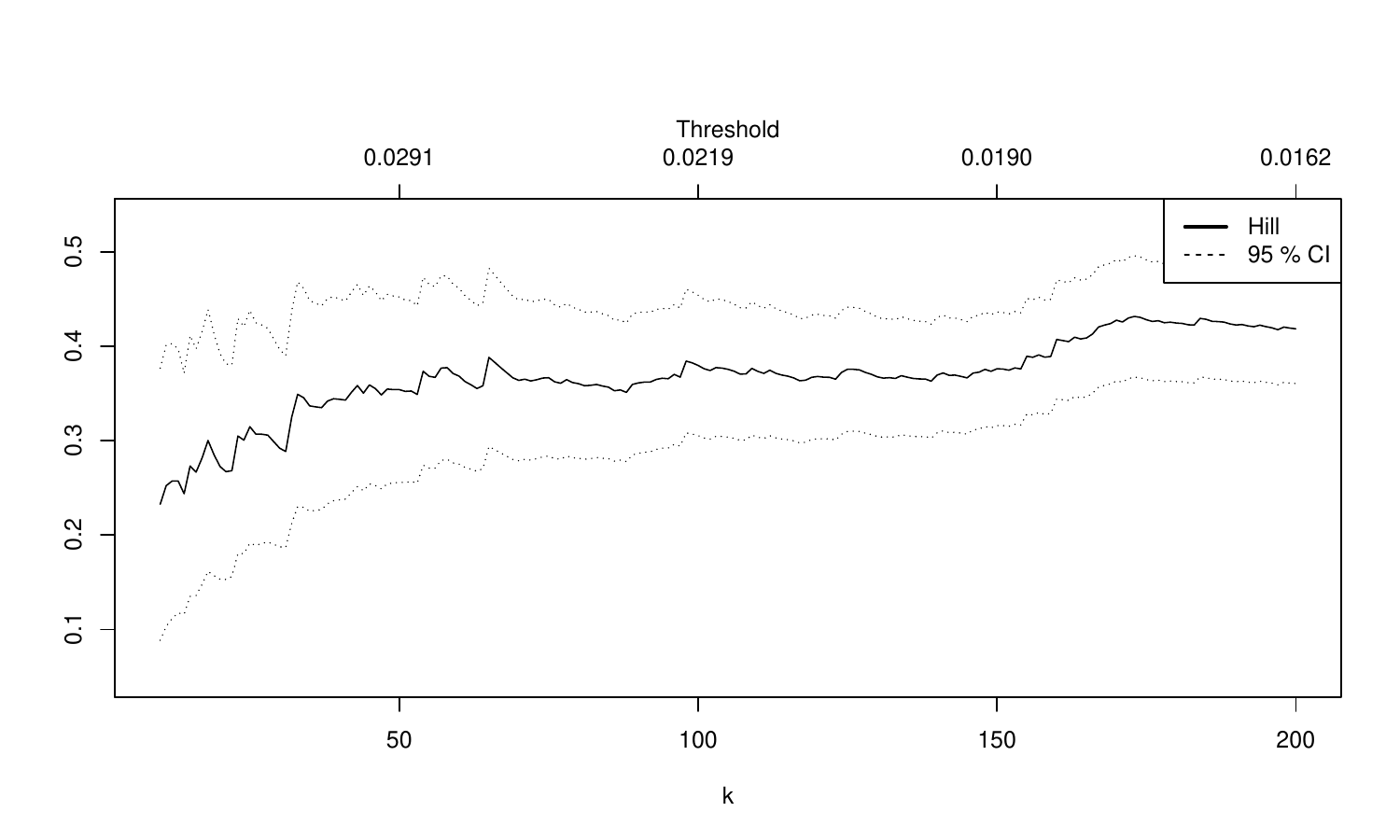}
    \caption{Hill estimates and the 95\% confidence interval of the extreme value index of $X_1$ for DJIA data.}
    \label{fig:hillplot}
\end{figure}

We choose the level of $k = 40$ by using a Hill plot of $X_1$, see Figure \ref{fig:hillplot}.
The Hill estimates of $\gamma_1$ and $\gamma_2$ are $0.344$ and $0.341$, both lie within the  confidence $D$.
This alignment supports the assumption that $\gamma_1 = \gamma_2$ is valid for this application.
Hill estimates derived from 30 individual institutions are detailed in  the Supplementary Material. Notably, only the estimates for   {\it Bank of America} and {\it The Procter \& Gamble} fall outside the confidence interval $D$, while the others remain within it. {For {\it Bank of America} and {\it The Procter \& Gamble}, we take a marginal transformation to $Y$ as follows. Define $\widetilde{Y} = |Y|^{\gamma_1/\gamma_Y}\text{sign}(Y)$ and $Y^* = \widetilde{Y}\text{sd}(Y)/\text{sd}(\widetilde{Y})$.  Such a transformation ensures that $Y^*$ shares the same extreme value index as $X_1$ while preserving the original scale of $Y$.
}

\begin{figure}[htb]
    \centering
    \includegraphics[width=0.8\textwidth]{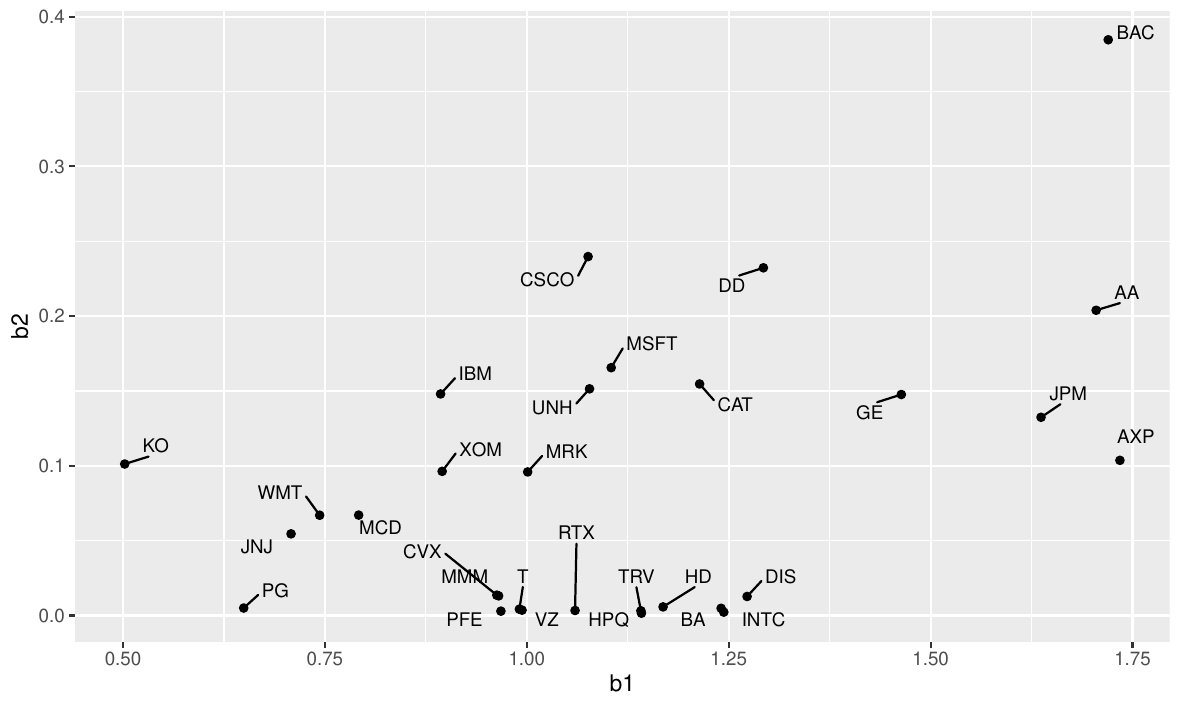}
    \caption{Scatter plot of  $\beta_1$ and $\beta_2$ for 30 companies from the DJIA.}
    \label{fig:scatterplot}
\end{figure}

We then employ the proposed estimator to estimate the coefficients $\beta_1$ and $\beta_2$ for all  the 30 institutions. The estimates are presented in  the Supplementary Material.
The scatter plot of
$\widehat{\beta}_1$ and $\widehat{\beta}_2$ for different institutions
are shown in Figure \ref{fig:scatterplot}.  {We observe the following patterns. 1) Regardless which tail-event occurred (systematic or systemic), BAC (Bank of America) suffered the highest impact, while institutions on the left bottom corner (PG (Procter \& Gamble), JNJ (Johnson \& Johnson), KO (Coca-Cola),  WMT (Walmart) and MCD (McDonald's)) belonging to consumer staple and health care industries received lowest impact, among all thirty institutions; 2) When a tail systematic event occurred, the institutions within the financial industry (BAC, JPM (JPMorgan Chase) and AXP (American Express)) and AA (Alcoa Corporation) suffered the highest impact; 3) When a tail-systemic event occurred, CSCO (Cisco), DD (DuPont de Nemours), AA  suffered the second highest impact.  These observations indicate  that financial institutions and manufacturing industry are expected to experience  higher losses compared to consumer staple and health care industry when the market crashes, when systematic and systemic events occurred, respectively.
These findings suggest that the proposed model can be utilized for tail-based clustering.}


\begin{figure}[htb]
    \centering
    \includegraphics[width=0.9\textwidth]{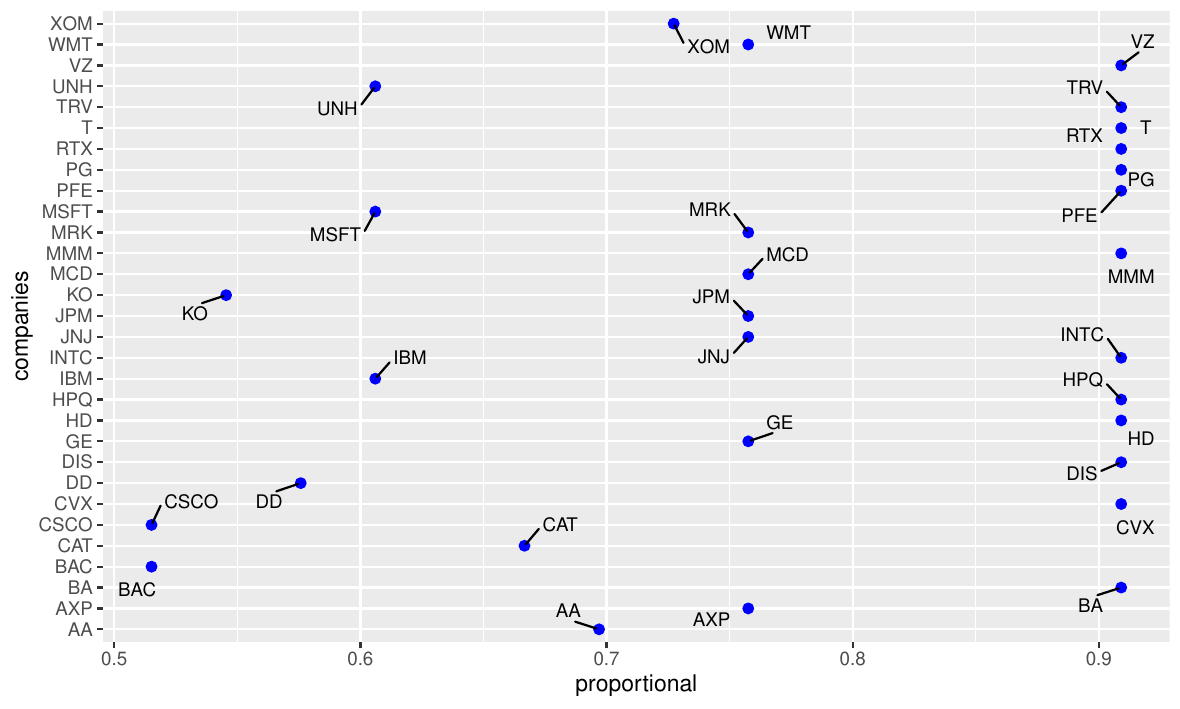}
    \caption{Scatter plot of   the proportion of days when $X_1$ dominates the risk of $Y$ for 30 companies from the DJIA.}
    \label{fig:propplot}
\end{figure}

We further analyze which factor dominates the tail risk of $Y$ when market crashes.  For this purpose, we compare the values of $\widehat{\beta}_1 X_1$ and $\widehat{\beta}_2X_2$ on the   system-wide stress events
$$
\mC = \set{i:X_{1i}> X_{(n-k^*),1} \ \text{or} \ X_{2i}> X_{(n-k^*),2} }.
$$
Here, we choose $k^* = 20$, a value smaller than $k$.  The choice  of $k^*$ reflects our belief that the tail regression model is valid only for a very small probability $p_0$. Nevertheless, using a relatively larger $k$ is useful  in the estimation process, as demonstrated by the simulation result.
The proportion of the days    when $\widehat{\beta}_1X_{1i}>\widehat{\beta}_2X_{2i}$ for $i\in  \mC$, denoted as $p_{\mC}$,  is presented in    the Supplementary Material and illustrated in the scatter plot in  Figure \ref{fig:propplot}. In general, $X_1$ has a higher impact on $Y$ compared to $X_2$ as  the value of $p_{\mC}$  are greater than 0.5 for all institutions. Additionally, systemic index ($X_2$) has a higher impact for BAC, CSCO and KO than other institutions.

\subsection{Rainfall data in Germany}
Beyond the financial market, we extend the application of our model to analyze daily rainfall data in Germany.
  In this analysis, $X_1$ represents the mean  of the rainfall amounts across multiple stations (systematic index), and $X_2$
  signifies the maximum of  the rainfall amounts among these stations (systemic index).

This dataset  encompasses daily rainfall amounts during summer seasons (from May to September),
recorded in 49 stations  in three regions of North–West Germany: Bremen, Nieder-sachsen and Hamburg. The dataset has been pre-processed by \cite{einmahl2022spatial} using declustering techniques
to reduce serial dependence.   The remaining datasets consists of $n=3552$ observations.   We define $X_1$ as the mean of the rainfall amounts across the stations and $X_2$ as the  cross sectional maximum of these rainfall amounts. Additionally, let $Y$ denote the daily rainfall amount of an individual institution.
This  dataset has been analyzed by \cite{einmahl2022spatial},
confirming that all the stations involved have the same positive extreme value indices.
The parameter $k$ is set at $60$, following analysis from a Hill plot of $X_1$ (figure omitted for brevity).
 The estimation results for $\beta_1$ and $\beta_2$ are  presented in  the Supplementary Material.

\begin{figure}[htb]
        \centering
        \subfigure[$\beta_1$]{
        \includegraphics[width=0.45\textwidth]{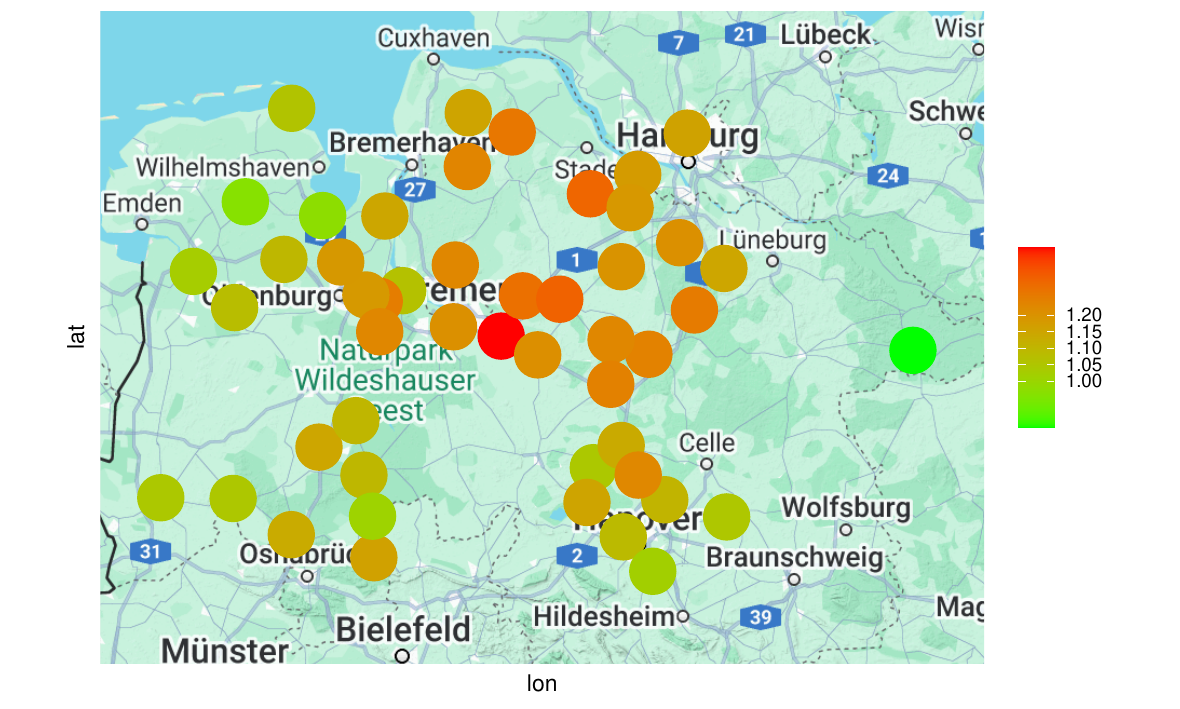}
        }
        \subfigure[$\beta_2$]{
            \includegraphics[width=0.45\textwidth]{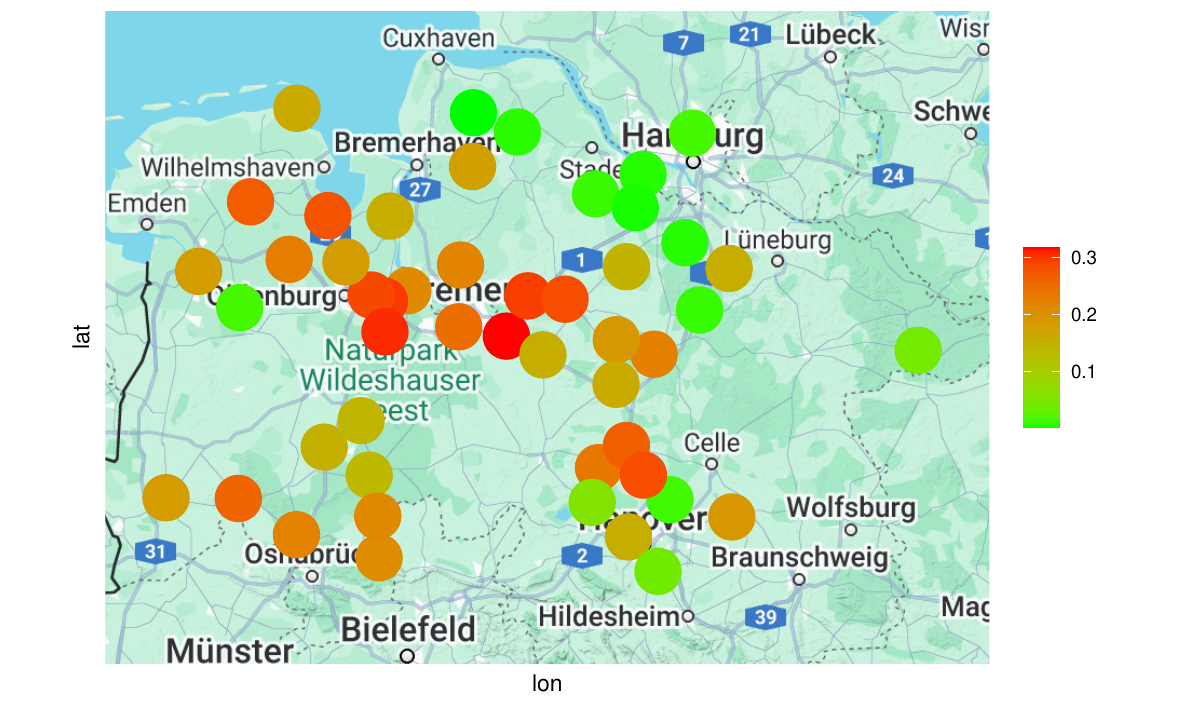}
        }
        \subfigure[$p_{\mC}$]{
            \includegraphics[width=0.45\textwidth]{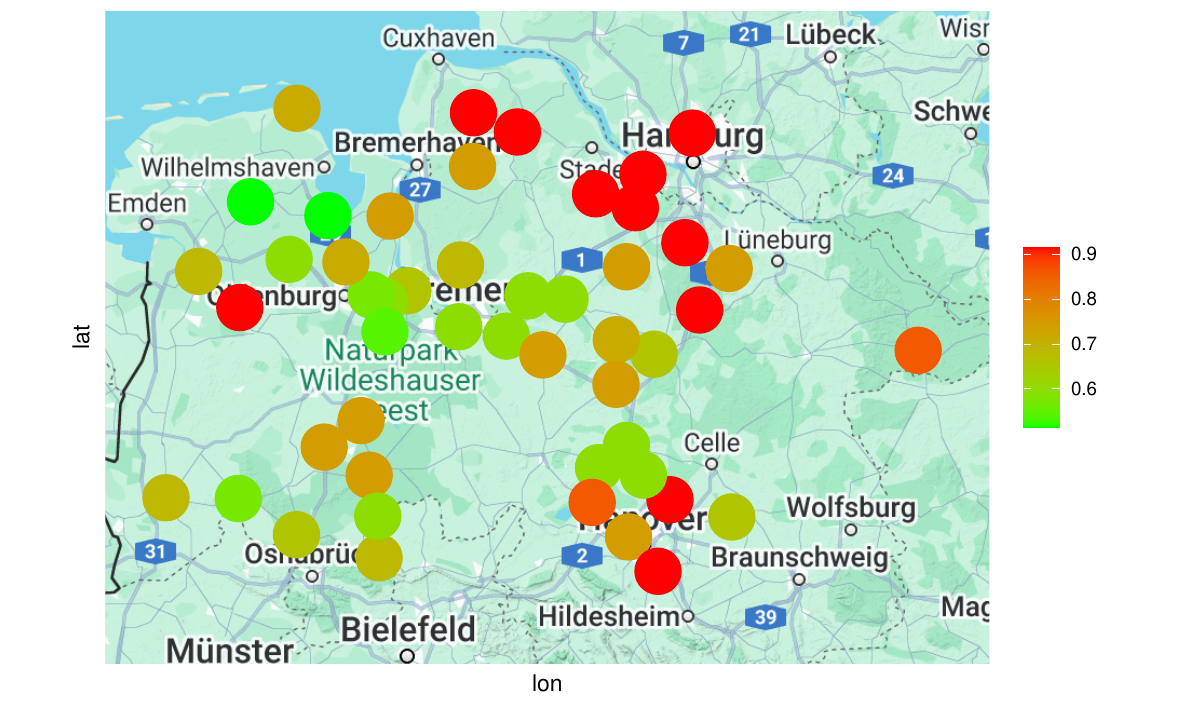}
            }
        \caption{Heat maps of $\beta_1$, $\beta_2$ and $p_{\mC}$ for all the 49 stations.}
        \label{fig:heat:gust}
    \end{figure}

    We present heat maps of $\beta_1$, $\beta_2$, and $p_{\mC}$ in Figure \ref{fig:heat:gust}, revealing that the systematic index ($X_1$) and systemic index ($X_2$) exert a greater influence on the response variable in the central region compared to other areas. Additionally, $p_{\mC}$ is lower in this central region, indicating a more significant contribution of the systemic index ($X_2$) to $Y$ in this region.

\section{Discussion}
In this paper, we introduced a max-linear tail regression model and applied it to two distinct datasets: the DJIA 30 stock returns, with predictors being the market index and the worst-case return scenarios, as well as rainfall data. Our model's novelty lies in its univariate treatment of individual components within the max-linear framework, which simplifies the analysis while capturing extreme behavior effectively.
Looking forward, we plan to enhance this model by extending the individual components to functions of multiple covariates, thereby enabling a more flexible and comprehensive analysis of complex systems with multiple interacting variables. This would allow us to better capture the nuances of covariate effects under extreme conditions and improve predictive accuracy across various domains.
Furthermore, while this paper focused on applications in financial markets and meteorological data, the methodology has broader implications. Future research can explore its potential in a wide range of applied sciences, including environmental modeling, engineering (e.g., stress analysis under extreme conditions), and risk management in insurance and economics. By adapting our framework to incorporate additional covariates and domain-specific nuances, this model has the potential to address critical challenges in understanding extreme events across diverse fields.

\FloatBarrier
\bibliographystyle{apalike}
\bibliography{mybib}
\end{document}